\documentclass[11pt]{article}
\usepackage{multicol}
\usepackage{graphicx}
\usepackage{amssymb,bm,mathrsfs,amscd}
\usepackage[tbtags]{amsmath}
\usepackage{indentfirst}
\begin{document}

\title{ The Roads to LPA Based Free Electron Laser }

\author{ Xiongwei Zhu \\
 Institute of High Energy Physics, \\
 Chinese Academy of Sciences, Beijing 100049 }

\maketitle

\begin{abstract}
   In this paper, we simply outline the present status of the free electron laser and the laser plasma based accelerator, and we
simply discuss the potential possible roads appearing in the accelerator community to use the laser plasma based accelerator into the field of the free electron laser.

\end{abstract}



\begin{Keywords}
 Laser Plasma Accelerator, Free Electron Laser, Compact Facility
\end{Keywords}
\section{Introduction}
    The conventional accelerator has the limit of the gradient due to the structure surface field limit, the best gradient achieved now is less than $ 100 MeV/m $.  Laser plasma accelerator ( LPA )\cite{1} will be the next generation accelerator with the ultrahigh gradient which will reach $ 100 GeV/m $.  Due to the ultrahigh gradient of LPA, the accelerator facility will be miniaturized greatly. So the linear collider and the free electron laser facility will be on the table-top. In case of the conventional accelerator, the electromagnetic wave in the accelerator structure is perfect or near perfect. So the produced electron beam is of high quality. While, the Langmuir wave excited in the plasma is not so perfect and not stable, so the produced electron beam quality is less perfect.

    Free electron laser\cite{2} is the coherent light source which needs the high quality electron beam. So the natural question is that it is possible to use the laser plasma accelerator into the free electron laser field. In this paper, we discuss this problem and give some possible roads.

\section{Free electron laser}
    Free electron laser is a kind of coherent radiation source and needs the high quality electron beam. In case of free electron laser, the wave particle resonant relation is
    \begin{equation}
        ( k_{z} + k_{w} )z - \omega t = 0,
    \end{equation}
    where $ k_{z} $ and $ k_{w} $ are the wavenumber of the pump wave and the undulator. The electrons resonate with the ponderomotive wave. From the resonant relation, the derived radiation wavelength is
    \begin{equation}
         \lambda = \frac{\lambda_{w}}{2 \gamma^2} ( 1 + \frac{K^2}{2} ),
    \end{equation}
    where $\lambda, \lambda_{w}$ are the wavelength of the radiation wave and the undulator period respectively, $\gamma$ is the relativistic factor of the electron beam, and $K$ is the strength of the undulator.  Due to the coherent requirement,
    we need the high quality beam. The beam emittance $\epsilon$ satisfies $\epsilon < \frac{\lambda}{4 \pi}$, and the beam relative energy spread satifies $\delta < \rho$, $\rho$ is the piece parameter.

    The high gain free electron laser\cite{3} was proposed by Bonifacio and Pellegrini, and was developed in the last twenties of years. The first lasing of the soft x-ray free electron laser occured in FLASH@DESY, while the first lasing of the hard x-ray free electron laser occured in LCLS@SLAC. There are many kinds of the operation modes for the short wavelength and high gain free electron laser, such as the SASE, HGHG, Self-seeding and EEHG. The frontier of the coherent x-ray light source is that the time resolution and the space resolution are $ 1 fs $, $ 0.1 nm $ respectively. As for the x-ray free electron laser, the typical required electron beam parameters are the beam charge of $ 1 pC-1 nC $, the energy of $ 1-10GeV $, the normalized emittance of $ 0.1-1 mm mrad $, the project energy spread of $ 0.01\%-0.1\% $, the peak current of $ 1-10kA $, and the output photon energy is $ 1-10keV $. The beam quality of the output electron beam of the present laser plasma accelerator can not reach this kind of parameters, so the laser plasma accelerator now can not be used as the free electron laser driver linac directly. In order to use LPA into FEL field\cite{4,5}, we should improve the output beam quality of LPA or we should find ways to compensate the relative energy spread.

\section{Laser plasma based accelerator}
    Since the first proposal of laser wakefield accelerator by Tajima and Dawson, the LPA has gone through over thirty years.  In laser plasma accelerator, the longitudinal Langmuir wave is excited by the laser which passes through the plasma. The
    perturbed electron density satifies the equation
    \begin{equation}
        ( \frac{\partial^2}{\partial t^2} + \omega_{p}^2 )\frac{\delta n}{n_{0}} = c^2 \nabla^2 \frac{a^2}{2},
    \end{equation}
    where $\delta n $ is the perturbed electron density, $ n_{0} $ is the unperturbed electron density, $\omega_{p}$ is the plasma frequency, $ a = \frac{e A}{m c^2} $ is the normalized laser strength, and $c$ is the light velocity. The maximal
    electric field is $ E= c m \omega_{p}/e  $. The perturbated electron density excites the plasma wave potential which forms the accelerating field as in conventional accelerator.

    With the appearance of the quasi-monoenergetic electron beam from laser plasma accelerator ( LPA ), there begins the new run of research interest. The produced beam quality of LPA depends on the injection method. The present main injection methods are the bubble mechanism, the colliding optical pulse injection, the controlling gradient method, and the external injection method. According to the laser strength, the laser plasma accelerator can be defined into three regimes: the linear regime, the quasilinear regime, and the nonlinear regime. In case of the linear regime or the linear theory, $ a_0 \ll 1$, the driving laser pulse will be guided well, and the excited plasma wave wakefield has fine consine structure of a few of cycles. While, in case of the nonlinear case or the blow-out regime, $ a_0 \gg 1$. The bubble mechanism works in this regime and can produce the quasi-monoenergytical electron beam. This kind of injection method work stably, but the wakefield wave form is not so perfect to go further to obtain the electron beam of the more better quality. In case of the quasilinear regime,  $ a_0 \sim 1$. This is the mid way to get the high quality electron beam. The quasilinear regime has the partial advantages of both the linear regime and the nonlinear regime. The bubble mechanism, the collidering optical pulse injection, the gradient control injection and the external injection method can work in this regime. In the past ten years, the bubble mechanism has made great breakthrough in the expedition to the advanced accelerator concept. The relative energy spread of LPA with the bubble mechanism now stay at the level of few or ten percent. It may be better to return to the case of relatively small $ a_0 $ to have the plasma wave wakefield of the perfect wave form/shape.

\section{ Roads to LPA based free electron laser}
    There are some experiments and proposals to use the laser plasma accelerator to drive free electron laser. The main obstacle for the application of laser plasma accelerator into free electron laser is the big energy spread of the produced electron beam of the laser plasma accelerator. In the early development of the laser plasma accelerator, the relative energy spread of the output electron beam is poor and completely $ 100\% $. But now, the best energy spread result of the laser plasma accelerator experiment is a few percents which still does not reach the idea value of $ \sim 0.1\% $. So there appear some methods to overcome the obstacle. These methods can be outlined mainly as following

    (1) The direct method is to use collimator to chop the output electron beam to get the required high quality electron beam for free electron laser application. In this way, we can arrange the chicane and slit to collimate the beam, as shown in Figure 1. Up to now, this method is the most practical way to use laser plasma accelerator as the FEL driver linac.

    \begin{center}
    \includegraphics[width=10 cm]{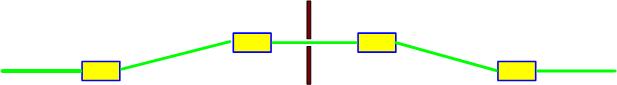}

    { Figure 1.The schematic figure of the collimating method. }
    \end{center}

    (2) Recently, Dr.Zhirong Huang from SLAC propose to use TGU\cite{6,7,8,9} ( transverse gradient undulator )theory to realize free electron laser by laser plasma accelerator. TGU theory uses the transverse magnetic field gradient to compensate the large energy spread. The transverse field gradient can be realized by canting undulator. TGU theory was first proposed by Smith, and Kroll\cite{8,9}.

    (3) In comparision with compression, Dr.Schroeder from LBNL propose to use the chicane to stretch the electron beam so as to reduce the relative energy spread, i.e. the decompresssion method\cite{10,11,12}. On the one hand, the energy spread of the electron beam is reduced to be fit to the free electron laser radiation process. On the other hand, the length of the electron beam is lengthened to mitigate the slippage effect. The difficulty of the decompression is that the output electron beam should has the good correlated energy spread to be fit to the decompression. But it is not so easy to control/tune the laser plasma accelerator now.

    (4) All optical free electron laser scheme\cite{13,14}. One use the plasma eigenmode or the laser as the pump wave ( RF undulator ) to produce the coherent radiation. Using RF undulator, it is convenient to produce short wavelength free electron laser. Further, we can use laser plasma scattering to produce x-ray radiation.

    (5) In order to make the LPA enter into the realistic application, we\cite{15} propose to first use laser plasma accelerator as the preinjector for the free electron laser driver linac at 2011 or we use LPA first as the electron source of the ultrahigh brightness. In this way, we use the main conventional RF linac to compensate the large energy spread to the required value. Further, due to the high brightness electron beam, we desert the usual bunch compressor technique. As an example, Figure 1 shows the basic scheme setup. This scheme reduce greatly the complexity of the FEL facility.

    \begin{center}
    \includegraphics[width=10 cm]{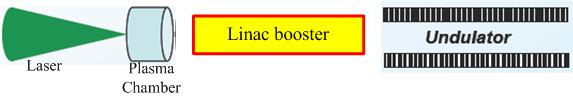}

    { Figure 2.The schematic layout of XFEL experiment. }
    \end{center}

\section {Discussion}
    In the past ten years, LPA has made great exciting progress towards the quasi-monoenergytic electron beam. But the relative energy spread of the produced electron beam still stays at the level of few percents which is still far bigger than the required value for x-ray free electron laser. There still exists long way to improve the beam quality of LPA to the ideal value. The bubble mechanism operates at the nonlinear regime which the excited Langmuir wave is not well perfect. The reason of the high beam quality in the conventional free electron laser facility is that the excited electromagnetic wave in the RF accelerating structure is so perfect as to produce the perfect electron beam to drive the free electron laser process. If one wants to get the perfect plasma wave in LPA, we may return to the original linear regime.

\end{document}